\documentclass[prd,twocolumn]{revtex4}
\usepackage{amsmath}
\usepackage{latexsym}
\usepackage{amsfonts}
\usepackage{epsf}
\begin{document}
\title{Observational constraints on generalized Chaplygin gas model}
\author{Yungui Gong} \email{gongyg@cqupt.edu.cn}
\affiliation{Institute of Applied Physics and College of
Electronic Engineering, Chongqing University of Posts and
Telecommunications, Chongqing 400065, China}
\begin{abstract}
The generalized Chaplygin gas model with parameter space
$\alpha>-1$ is studied in this paper. Some reasonable physical
constraints are added to justify the use of the larger parameter
space. The Type Ia supernova data and age data of some clusters
are then used to fit the model. We find that the parameters have
bimodal distributions. For the generalized Chaplygin gas model, we
also find that less free parameters fit the data better. The best
fit model is the spatially flat model with baryons. The best fit
parameters are: $\Omega_{m0}=0.044$, $w_{c0}=-0.881$ and
$\alpha=1.57$. The transition redshift is $z_T=0.395$.
\end{abstract}
\maketitle

\section{Introduction}
The mounting  evidences suggest that the Universe is expanding
with acceleration \cite{sp99,agr04,wmap}. This reveals that the
Universe is dominated by dark energy with negative pressure, whose
energy density fraction is about 2/3 at present. The nature and
origin of dark energy is still mysterious, leaving theoretic
physicts searching for novel answers. Althoug a cosmological
constant may be responsible for the accelerating expansion of the
Universe and be consistent with current observational data, the
extremely smallness of the value of the cosmological constant
brings a big challenge to us. It is also possible that the
mechanism at work is dynamical. A dynnamical candidate to provide
dark energy may be supplied by a slowly rolling scalar field,
widely referred as ``quintessence field'' \cite{quint}. There are
also models of a scalar field with non-canonical kinetic term,
known as ``k-essence'' \cite{kessence} or ``tachyonic'' models
\cite{tachyon}. Other alternative dark energy models include the
holographic dark energy model \cite{holo} and the
extra-dimensional motivated models \cite{dgp}.

Among the cosmological community, the consensus about our Universe
is the so called ``concondance model'': 95\% percent of the
Universe is composed of dark components. Recently the Chaplygin
gas model with exotic equation of state $p=-A/\rho$ was proposed
to explain both dark enrgy and dark matter \cite{chap}. The
Chaplygin gas model was later generalzied to the generalized
Chaplygin gas (GCG) model with equation of state
$p_{c}=-A/\rho_{c}^\alpha$ \cite{gcg}. The novel feature of this
model is that it unifies dark energy and dark matter in one model.
Sandvik and coauthors claimed that the matter power spectrum
essentially ruled the GCG model out \cite{sandvik}. However, their
analysis does not include the effect of the baryons. Be\c{c}a and
collabrators showed that it is important to include baryons in the
study of large scale structure and the conclusion changed when
baryons were included \cite{beca}. It should also be noted that
the results in \cite{sandvik} was based on linear theory of
perurbations and neglected any non-linear effects.

The GCG model has been extensively studied in the literature
\cite{gcg1,gcg2}. The GCG model was also studied in the framework
of modified Friedman equation in \cite{gcgmfe}. In this paper, we
consider both dust like matter and GCG as the source. The dust
like matter may be just baryons or a portion of dark matter.
Unlike other studies, we allow the parameter $w_{c0}=-A_s$ to be
less than $-1$ and the parameter $\alpha$ to be in the region
$-1<\alpha<0$ in addition to the usual region $\alpha>0$. Some
physically reasonable conditions are then applied to the model to
constrain the parameters. Therefore, the model considered here is
more general in addition to be physical.
\section{GCG model}
In a homogeneous and isotropic universe, the
Friedmann-Robertson-Walker (FRW) space-time metric is
\begin{equation}
\label{rwmetric} ds^2=-dt^2+a^2(t)\left[{dr^2\over
1-k\,r^2}+r^2\,d\Omega\right].
\end{equation}
For a null geodesic, we have \begin{equation} \label{line}
\int_{t_1}^{t_0}{dt\over a(t)}=\int_0^{r_1}{dr\over
\sqrt{1-kr^2}}\equiv f(r_1), \end{equation} where
\begin{equation*}
f(r_1)=\left\{\begin{array}{ll}
\sin^{-1}r_1,\ \ \ \ \ \ &k=1,\\
r_1,&k=0,\\
\sinh^{-1}r_1,&k=-1.
\end{array}\right.
\end{equation*}
With both an ordinary pressureless dust matter and GCG as sources,
the Friedmann equations read
\begin{gather}
\label{cos1} H^2+{k\over a^2}={8\pi
G\over 3}(\rho_m+\rho_c),\\
\label{cos2} \dot{\rho_c}+3H(\rho_c+p_c)=0,
\end{gather}
where the Hubble parameter $H=\dot{a}/a$, dot means derivative
with respect to time, $\rho_m=\rho_{m0}(a_0/a)^3$ is the matter
energy density, a subscript 0 means the value of the variable at
present time. By using the GCG equation of state
$p_{c}=-A/\rho_{c}^\alpha$, we get the solution to Eq.
(\ref{cos2}) as \begin{equation} \label{eqrhoch}
\rho_{c}=\left[A+{B\over a^{3(1+\alpha)}}\right]^{1/(1+\alpha)}.
\end{equation} Because $w_{c}=p_c/\rho_c=-A/\rho^{\alpha+1}_c$, so
$A=-w_{c0}\rho^{\alpha+1}_{c0}$. Substitute this expression into
Eq. (\ref{eqrhoch}), we get
$B=(1+w_{c0})a^{3(1+\alpha)}_0\rho^{\alpha+1}_{c0}$. Therefore,
Eq. (\ref{eqrhoch}) can be expressed in terms of $w_{c0}$ and
$\rho_{c0}$ as \begin{equation} \label{eqrhoch1}
\rho_{c}=\rho_{c0}\left[-w_{c0}+(1+w_{c0})\left({a_0\over
a}\right)^{3(1+\alpha)}\right]^{1/(1+\alpha)}. \end{equation} It
is obvious that the generalized Chaplygin gas behaves like the
cosmological constant when $w_{c0}=-1$ and it behaves like the
dust matter when $w_{c0}=0$. At early times, i.e., the
cosmological radius $a(t)$ is small, $\rho_c\sim (a_0/a)^3$, which
corresponds to a dust like dominated universe. At late times,
i.e., the cosmological radius $a(t)$ is large, $\rho_c\sim {\rm
constant}$, which corresponds to a cosmological constant like
dominated universe. Therefore the generalized Chaplygin gas
interpolates between a dust dominated phase in the past and a
de-Sitter phase at late times. This distinct feature makes the
model an intriguing candidate for the unification of dark matter
and dark energy.

The GCG equation of state can be derived from the generalized
Born-Infeld action \cite{bento} \begin{equation}
\mathcal{L}=-A^{1/(1+\alpha)}\left[1-(g^{\mu\nu}\phi,_\mu\phi,_\nu)^{(1
+\alpha)/2\alpha}\right]^{\alpha/(1+\alpha)},\end{equation} where
$\phi,_\mu=\partial \phi/\partial x^\mu$. From the above
Lagrangian, we can easily get $p_c=\mathcal{L}=-A/\rho^\alpha_c$.
If we take GCG as a quintessence field, i.e., if we take
$w_{c0}\ge -1$, then if GCG is the only source in a spatially flat
universe, the potential for GCG is \begin{equation}\begin{split}
V(\phi)={A^{1/(1+\alpha)}\over
2}\left[\cosh^{-2\alpha/(1+\alpha)}\left({3(1+\alpha)\phi\over
2}\right) \right. \\ \left.
+\cosh^{2/(1+\alpha)}\left({3(1+\alpha)\phi\over 2}\right)\right],
\end{split}\end{equation} here we set $8\pi G=3$.

 By using Eq. (\ref{eqrhoch1}), we get \begin{equation}\label{eqpch}
p_c=w_{c0}\rho_{c0}\left[-w_{c0}+(1+w_{c0})\left({a_0\over
a}\right)^{3(1+\alpha)}\right]^{-\alpha/(1+\alpha)}.
\end{equation} If $a_0/a\ll 1$, then to the first order of
expansion, $\rho_c$ and $p_c$ are \begin{equation}
\rho_c=\rho_{c0}(-w_{c0})^{1/(1+\alpha)}\left[1-{1+w_{c0}\over
w_{c0}(1+\alpha)}\left({a_0\over
a}\right)^{3(1+\alpha)}+\cdots\right],\end{equation}
\begin{equation}
p_c=-\rho_{c0}(-w_{c0})^{1/(1+\alpha)}\left[1+{(1+w_{c0})\alpha\over
w_{c0}(1+\alpha)}\left({a_0\over
a}\right)^{3(1+\alpha)}+\cdots\right].\end{equation} From the
above expressions, we see a mixture of a cosmological constant
with a type of dark energy described by a constant equation of
state parameter $\alpha$. So the physical meaning of $\alpha$ may
be given in this sense.

Follow Chiba and Nakamura \cite{chiba}, we require the following
physically reasonable conditions: (1) The current total energy
density is non-negative; (2) The total energy density is not
increasing; (3) The present sound speed $c_s$ of the system
satisfies $0\le c_s^2\le 1$ because of causality and local
stability. The first condition gives $\Omega_k\le 1$, where
$\Omega_k=-k/(a^2_0H^2_0)$. The second condition tells us that
$1+q_0-\Omega_k\ge 0$, where the deceleration parameter
$q=-\ddot{a}/(aH^2)$. The second condition is equivalent to the
requirement that the effective equation of state for the total
source $w_{eff}\ge -1$. The third condition gives $1-\Omega_k\le
j_0\le 4(1-\Omega_k)+3q_0$, where the jerk parameter
$j=\dddot{a}/(aH^3)$. The deceleration parameter $q_0$ and the
jerk parameter $j_0$ are similar to the state finder parameters
used in \cite{finder}.

For the generalized Chaplygin gas model, we get
\begin{equation}
\label{eqj0} j_0=1-\Omega_k-{9\over 2}\alpha
w_{c0}(1+w_{c0})\Omega_{c0},
\end{equation}
\begin{equation}
\label{eqq0} q_0=-1+{3\over 2}\Omega_{m0}+\Omega_k+{3\over
2}(1+w_{c0})\Omega_{c0},
\end{equation}
where $\Omega_m\{\Omega_c\}=8\pi G\rho_m\{\rho_c\}/(3H_0^2)$ and
$\Omega_{c0}=1-\Omega_{m0}-\Omega_k$. By using the above Eqs.
(\ref{eqj0}) and (\ref{eqq0}), we get the following constraints
\begin{equation}
\label{cond1}
 \Omega_{m0}+(1+w_{c0})\Omega_{c0}\ge 0,
\end{equation}
\begin{equation}
\label{cond2} \alpha w_{c0}(1+w_{c0})\Omega_{c0}\le 0,
\end{equation}
\begin{equation}
\label{cond3} \Omega_{m0}+(1+w_{c0})(1+\alpha
w_{c0})\Omega_{c0}\ge 0.
\end{equation}
Furthermore, we require that $\Omega_{m0}\ge 0$ and
$\Omega_{c0}\ge 0$. To get accelerated expansion, we also require
that $q_0< 0$. Therefore, we have one additional constraint
\begin{equation} \label{cond4} \Omega_{m0}+(1+w_{c0})\Omega_{c0}<
{2\over 3}(1-\Omega_k). \end{equation} In the literature, the
parameters are usually constrained to be $w_{c0}\ge -1$ and
$0<\alpha\le 1$. Some authors also considered the possibility of
$\alpha>1$. From Eqs. (\ref{cond1}-\ref{cond4}), we see that it is
possible to get $w_{c0}< -1$ if $\alpha<0$. In this paper, we
consider the parameter space to be: $\Omega_{m0}=[0,1]$,
$w_{c0}=[-3,0)$ and $w_{c0}\neq -1$, $\alpha=(-1,100]$. In
addition to the constraints (\ref{cond1}-\ref{cond4}) on the
parameters, we also require that the energy density of GCG given
in Eq. (\ref{eqrhoch1}) is not negative.

Combining Eqs. (\ref{cos1}), (\ref{cos2}) and (\ref{eqrhoch1}), we
get the transition redshift $z_T$ when the expansion of the
Universe underwent the transition from deceleration to
acceleration \begin{equation} \label{eqzt}
\begin{split}\frac{\Omega_{m0}}{\Omega_{c0}}\left[-w_{c0}+(1+w_{c0})(1+z_T)^{3(1+\alpha)}\right]^{\alpha/(1+\alpha)}\\
=-2w_{c0}(1+z_T)^{-3}-(1+w_{c0})(1+z_T)^{3\alpha}.\end{split}\end{equation}
\section{Supernova Ia Fitting Results}
In this section, we use the 157 gold sample supernova Ia (SN Ia)
data compiled in \cite{agr04} to fit the model. The parameters in
the model are determined by minimizing
\begin{equation}\label{chi2} \chi^2=\sum_i{[\mu_{\rm
obs}(z_i)-\mu(z_i)]^2\over \sigma^2_i},\end{equation} where the
extinction-corrected distance moduli $\mu(z)=5\log_{10}(d_{\rm
L}(z)/{\rm Mpc})+25$, the redshift $z=a_0/a-1$, the luminosity
distance is \begin{equation} \begin{split} d_{\rm L}
=a_0(1+z)r_1=a_0(1+z){\rm sinn} \left[\frac{1}{a_0H_0}\int_0^z
\frac{dz'}{E(z')}\right]\\
=\frac{1+z}{H_0\sqrt{|\Omega_k|}} {\rm
sinn}\left[\sqrt{|\Omega_k|}\int_0^z
\frac{dz'}{E(z')}\right],\end{split} \end{equation} the
dimensionless Hubble parameter
$E(z)=H(z)/H_0=\Omega_m+\Omega_c+\Omega_k(1+z)^2$, ${\rm sinn}(x)$
is defined as $\sin(x)$\{$x$, $\sinh(x)$\} if $k=1$\{$0$, $-1$\}
respectively, and $\sigma_i$ is the total uncertainty in the
observation. The nuisance parameter $H_0$ is marginalized over
with a flat prior assumption. Since $H_0$ appears linearly in the
form of $5\log_{10}H_0$ in $\chi^2$, so the marginalization by
integrating $L=\exp(-\chi^2/2)$ over all possible values of $H_0$
is equivalent to finding the value of $H_0$ which minimizes
$\chi^2$ if we also include the suitable integration constant.
Because we assume a flat prior on $H_0$, therefore alternatively
we marginalize $H_0$ by minimizing $\chi^{\prime
2}=\chi^2(y)-2\ln(10)\,y/5-2\ln[\ln(10)\sqrt{(2\pi/\sum_i
1/\sigma^2_i)}/5]$ over $y$, where $y=5\log_{10}H_0$. To get the
marginalized likelihood of a parameter, we marginalize all other
parameters by integrating the probability distribution
$L=\exp(-\chi^2/2)$ over all possible values of the other
parameters.

\begin{figure}[htb]
\vspace{-0.1in} \epsfxsize=3in\epsffile{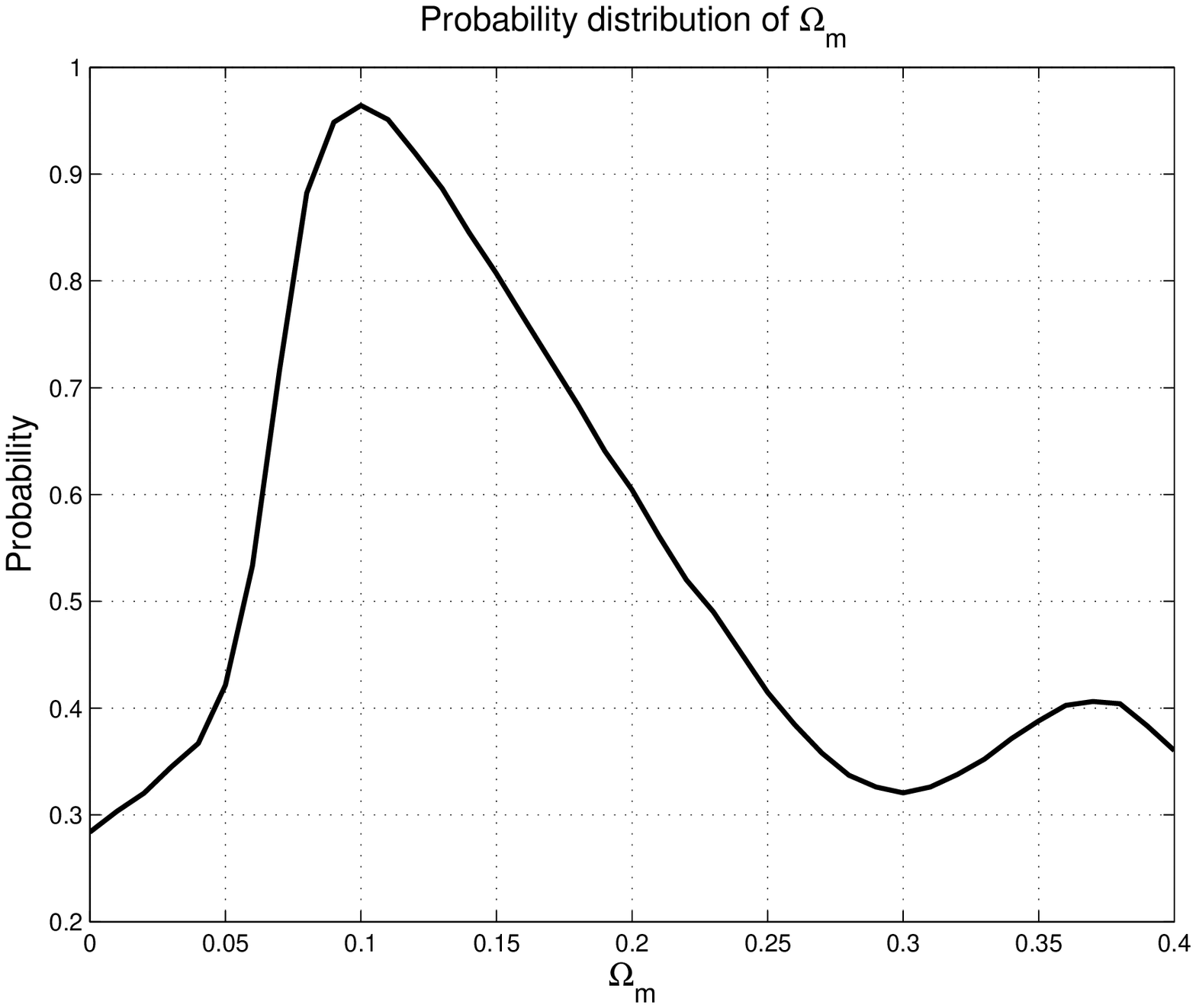} \vspace{-0.1in}
\caption{The marginalized likelihood distribution of $\Omega_{m0}$
for the spatially flat GCG model fitting result to SN Ia data.}
\label{wprob}
\end{figure}

\begin{figure}[htb]
\vspace{-0.1in} \epsfxsize=3in\epsffile{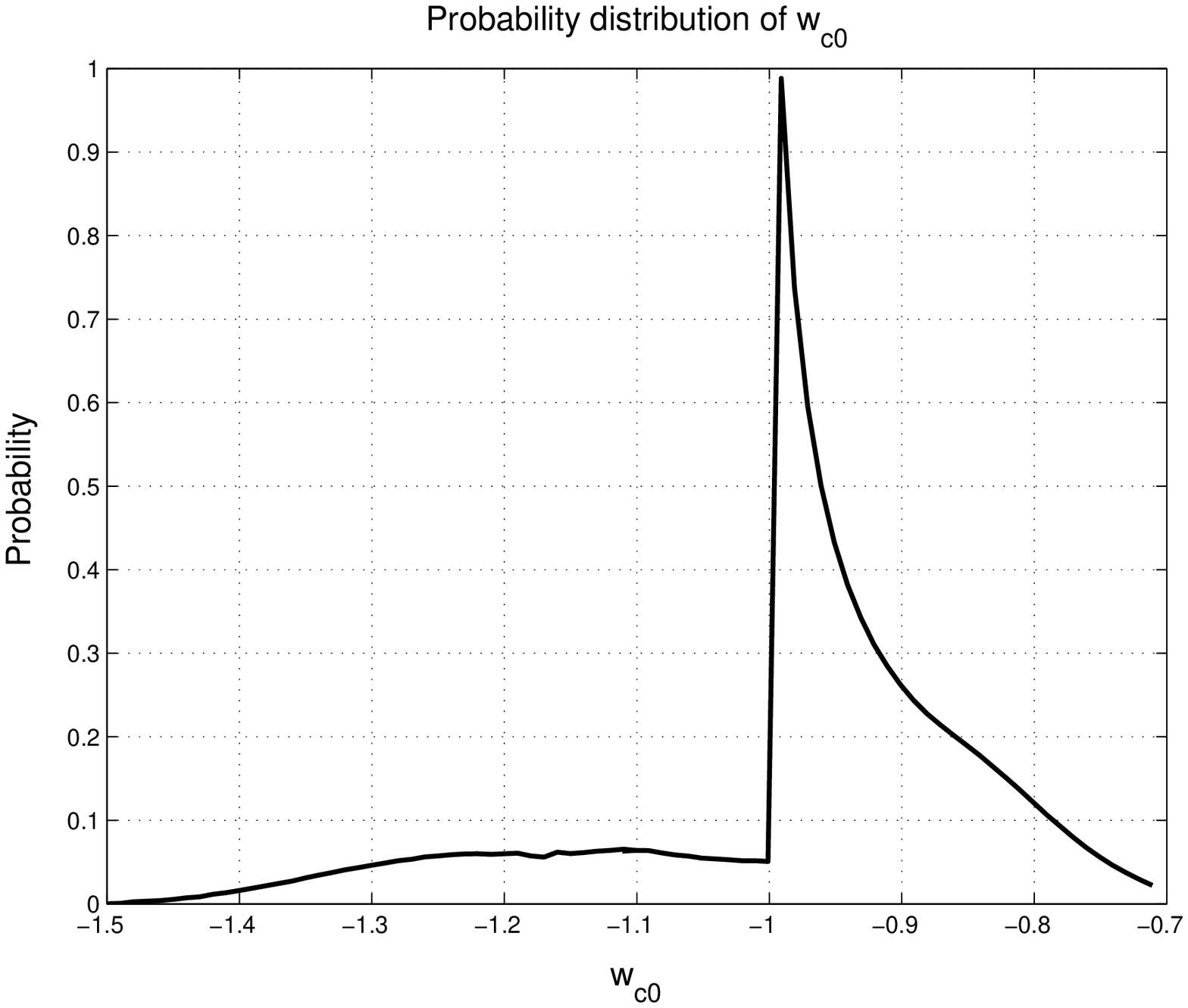} \vspace{-0.1in}
\caption{The marginalized likelihood distribution of $w_{c0}$ for
the spatially flat GCG model fitting result to SN Ia data.}
\label{cprob}
\end{figure}

\begin{figure}[htb]
\vspace{-0.1in} \epsfxsize=3in\epsffile{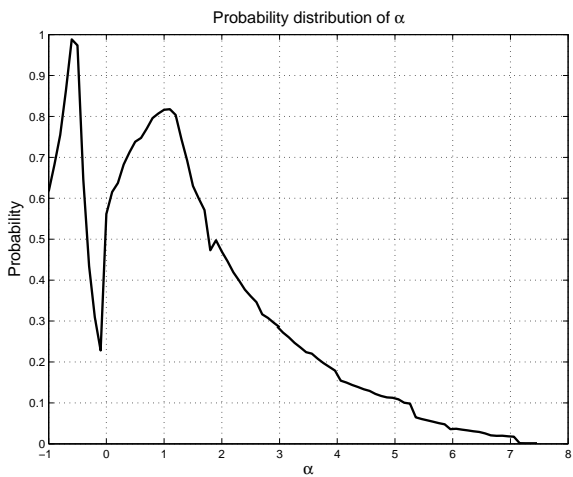} \vspace{-0.1in}
\caption{The marginalized likelihood distribution of $\alpha$ for
the spatially flat GCG model fitting result to SN Ia data.}
\label{aprob}
\end{figure}

We first consider the special case $w_{c0}=-1$, i.e., the LCDM
(Lambda cold dark matter) model. The best fit to the Riess gold
data is $\Omega_{m0}=0.31\pm 0.04$ with $\chi^2=176.5$. The Akaike
information criterion (AIC) is $\chi^2+2=178.5$ and the Bayesian
information criterion (BIC) is $\chi^2+\ln(157)=181.56$
\cite{stats}. Next we consider the spatially flat case
$\Omega_k=0$, the best global fit gives that
$\Omega_{m0}=0.073^{+0.20}_{-0.015}$,
$w_{c0}=-0.97^{+0.18}_{-0.03}$ and $\alpha=3.42^{+2.57}_{-2.67}$
with $\chi^2=173.66$. Due to the constraints
(\ref{cond1}-\ref{cond4}), it is difficult to find contours for
the parameters. The AIC is $\chi^2+2\times 3=179.66$ and the BIC
is $\chi^2+3\ln(157)=188.83$. For $w_{c0}<-1$ and $\alpha<0$, We
get the local best fit: $\Omega_{m0}=0.43^{+0.04}_{-0.07}$,
$w_{c0}=-1.40^{+0.18}_{-0.08}$ and $\alpha=-0.55^{+0.09}_{-0.04}$
with $\chi^2=174.12$. The AIC is $\chi^2+2\times 3=180.12$ and the
BIC is $\chi^2+3\ln(157)=189.29$. The probability distributions of
$\Omega_{m0}$, $w_{c0}$ and $\alpha$ are shown in Figs.
\ref{wprob}-\ref{aprob}. It is obvious that the distributions have
bimodal characteristics. Therefore the models $p_c=-A/\rho^\alpha$
and $p=-A\rho^\alpha$ fit the supernova data almost equally well.
The best fit of the full model gives that $\Omega_{m0}=0.0025$,
$\Omega_k=0.23$, $w_{c0}=-0.9997$ and $\alpha=10.3$ with
$\chi^2=173.46$. Apparently, the full GCG model with curvature
term fits worse than the spatially flat GCG model and the full
model tends to be curved LCDM model.

If we think that the dust like matter source consists baryons
only, then we can add a prior $\Omega_{m0}=0.044\pm 0.004$
\cite{wmap1} on the GCG model. In this case, for a spatially flat
universe, the best fit parameters are $\Omega_{m0}=0.044$,
$w_{c0}=-0.88^{+0.08}_{-0.03}$ and $\alpha=1.57^{+0.1}_{-0.94}$
with $\chi^2=173.95$. Substituting the best fit parameters into
Eq. (\ref{eqzt}), we find the transition redshift is $z_T=0.395$.
The above results are summarized in Table \ref{tab1}. Therefore,
the current SN Ia data does not favor GCG model over LCDM model.
Furthermore, more parameters fail to give better fit.
\begin{table}[htp]
\begin{tabular}{|l|c|c|c|} \hline
Model&$\chi^2$&AIC&BIC\\\hline LCDM&176.5&178.5&181.56\\\hline
GCG1&173.66&179.66&188.83\\\hline
GCG2&174.12&180.12&189.29\\\hline
GCG3&173.95&177.95&184.06\\\hline
\end{tabular}
 \caption{The comparison between different models. GCG1 refers to the
 globally best fit spatially flat GCG model, GCG2 refers to the locally best fit spatially
 flat GCG model and GCG3 refers to the spatially flat GCG model with the assumption that
 the dust like matter source is only baryons.} \label{tab1}
\end{table}

\section{Age Fitting Results}
Follow \cite{age}, we define the look back time as
\begin{equation} \label{tlz} t_L(z)={1\over H_0}\int_0^z
\frac{dx}{(1+x)E(x)}. \end{equation} The current age of the
Universe is $t_0=t_L(\infty)=14.4\pm 1.4$ Gyr \cite{age,age1}. The
age of an object $i$ at redshift $z$ is given by \begin{equation}
\label{tcl} t_i(z)={1\over H_0}\int_z^{z_F}
\frac{dx}{(1+x)E(x)}=t_L(z_F)-t_L(z), \end{equation} where $z_F$
is the formation redshift when the object was born. From Eq.
(\ref{tcl}), we see that
$$t_L(z)=t_L(z_F)-t_i(z)=t_0-t_i(z)-[t_0-t_L(z_F)],$$
where the delay factor $df=t_0-t_L(z_F)$ gives the information
about the unknown formation redshift $z_F$. As in \cite{age}, we
also assume that the delay factor is the same for all the objects.
The parameters in the GCG model are determined by minimizing
\begin{equation}\label{chi1} \tilde{\chi}^2=\left(\frac{t_0-t^{\rm
obs}_0}{\sigma_t}\right)^2+\sum_i\frac{[t_L(z_i)-t_0^{\rm
obs}+t_i^{\rm
obs}(z_i)+df]^2}{\sigma^2_t+\sigma_c^2},\end{equation} where
$t_0^{\rm obs}=14.4$ Gyr, $\sigma_t=1.4$ Gyr and $\sigma_c=1$ Gyr
\cite{age}. The nuisance parameter $df$ is marginalized over by
integrating the likelihood function $L=\exp(-\tilde{\chi}^2/2)$
over all possible values of $df$. Alternatively, we marginalize
over $df$ by minimizing $\tilde{\chi}^2$ over $df$ which gives
$df=-\sum_i^n [t_L(z_i)-t^{\rm obs}_0+t^{\rm obs}_i(z_i)]/n$.
Because we already have four parameters: $\Omega_{m0}$,
$\Omega_k$, $w_{c0}$ and $\alpha$ in the model, we use
$H_0^{-1}=9.78h^{-1}$ Gyr with $h=0.72$ given by HST Key project
\cite{hst}. The observational data for the age of cluster sample
is given in \cite{age}. We reproduce the data in Table \ref{tab2}.
\begin{table}[htp]
\begin{tabular}{|l|c|c|c|c|c|c|} \hline
$z$&0.1&0.25&0.6&0.7&0.8&1.27\\\hline $t_i(z)$
(Gyr)&10.65&8.89&4.53&3.93&3.41&1.6\\\hline $t_0^{\rm obs}-t_i(z)$
(Gyr)&3.75&5.51&9.87&10.47&10.99&12.8\\\hline
\end{tabular}
 \caption{The age data from Ref. \cite{age}} \label{tab2}
\end{table}

For the spatially flat LCDM model, the best fit result is
$\Omega_{m0}=0.20^{+0.08}_{-0.06}$ with $\chi^2=1.0$. Due to the
sparse of the data, the result is not as good as that from SN Ia.
For the spatially flat GCG model, the best fit results are:
$\Omega_{m0}\sim 0$, $w_c\sim -1$ and $\alpha=6.53$. Again, the
data does not favor GCG model over LCDM model.

\section{Conclusions}
In this paper, we studied the GCG model which provides a
phenomenological mechanism of unifying dark matter and dark
energy. We explored a larger parameter space for the GCG model.
Instead of studying the usual parameter range $0\le\alpha\le 1$,
we extended the parameter space to be $\alpha>-1$ with some
physical constraints on the parameters. We found that the
parameters had bimodal distributions in general. The constraints
from Type Ia SN data and the age data of clusters do not favor the
GCG model over the simplest LCDM model from the standards of the
Akaike information criterion and the Bayesian information
criterion. Therefore, the current observations are consistent with
both the GCG model and the LCDM model. Moreover, the flat model
fits the observational data better. The only benefit of the GCG
model is that it may provide a phenomenological mechanism of
unifying dark matter and dark energy.

\begin{acknowledgments}
The author YG thanks the hospitality of the Abdus Salam
International Center of Theoretical Physics where part of the work
was done. The author YG is also thankful to the ICTS of the
University of Science and Technology of China for organizing the
String/M theory, particle physics and Cosmology workshop where
this work was presented and discussed. The work is fully supported
by CQUPT under grants A2003-54 and A2004-05.
\end{acknowledgments}

\end{document}